\documentclass{article}

\usepackage{arxiv}

\usepackage[utf8]{inputenc} % allow utf-8 input
\usepackage[T1]{fontenc}    % use 8-bit T1 fonts
\usepackage{hyperref}       % hyperlinks
\usepackage{url}            % simple URL typesetting
\usepackage{booktabs}       % professional-quality tables
\usepackage{nicefrac}       % compact symbols for 1/2, etc.
\usepackage{microtype}      % microtypography
\usepackage{lipsum}
\usepackage{graphicx}
\graphicspath{{./images/}}
\usepackage{hyperref}
\hypersetup{pdftex,colorlinks=true,allcolors=blue}
\usepackage{amsmath,amssymb,amsfonts}
\usepackage{stfloats}
\usepackage[]{algorithm2e}
\usepackage{siunitx}
\newcolumntype{C}{>{\centering\arraybackslash}p{1.75cm}}

\title{Wavelet Augmented Regression Profiling (WARP): improved long-term estimation of travel time series with recurrent congestion \thanks{This paper has been accepted to the IEEE 23rd International Conference on Intelligent Transporatation Systems (IEEE ITSC 2020) and was part funded by the EPSRC under grant no. EP/L015374.}}

\author{
  Alvaro Cabrejas Egea \thanks{Alvaro Cabrejas Egea is with Mathematics for Real-World Systems Centre for Doctoral Training, University of Warwick.} \\
  MathSys Centre for Doctoral Training, \\
  University of Warwick \& Alan Turing Institute\\
  CV4 7AL Coventry, NW1 2DB London, UK \\
  \texttt{a.cabrejas-egea@Warwick.ac.uk} \\
   \And
 Colm Connaughton \thanks{Colm Connaughton is with Warwick Mathematics Institute, University of Warwick.}\\
  Warwick Mathematics Institute\\
  University of Warwick\\
  CV4 7AL Coventry, UK \\
  \texttt{c.p.connaughton@warwick.ac.uk} \\
}

\begin{document}
\maketitle

\begin{abstract}
Reliable estimates of typical travel times allow road users to forward plan journeys to minimise travel time, potentially increasing overall system efficiency.
On busy highways, however, congestion events can cause large, short-term spikes in travel time. 
These spikes make direct forecasting of travel time using standard time series models difficult on the timescales of hours to days that are relevant to forward planning. 
The problem is that some such spikes  are caused by unpredictable incidents and should be filtered out, whereas others are caused by recurrent peaks in demand and should be factored into estimates. 
Here we present the Wavelet Augmented Regression Profiling (WARP) method for long-term estimation of typical travel times. 
WARP linearly decomposes historical time series of travel times into two components: background and spikes. 
It then further separates the spikes into contributions from recurrent and residual congestion. 
This is achieved using a combination of wavelet transforms, spectral filtering and locally weighted regression.
The background and recurrent congestion contributions are then used to estimate typical travel times with horizon of one week in an accurate and computationally inexpensive manner. 
We train and test WARP on the M6 and M11 motorways in the United Kingdom using 12 weeks of link level travel time data obtained from the UK's National Traffic Information Service (NTIS).
In out-of-sample validation tests, WARP compares favourably to estimates produced by a simple segmentation method and to the estimates published by NTIS.

\end{abstract}

\section{Introduction} \label{Introduction}
\subsection{Background} \label{Background}
In the United Kingdom, Highways England (HE) is responsible for most motorways and major roads in England, comprising the Strategic Road Network (SRN).
The SRN is monitored by the National Traffic Information System \cite{NTIS}, which gathers and processes travel time, speed, flow and headway data in real time using road sensors and special vehicles.
The smallest components of the SRN are the "links", segments of motorway ranging from 500 to 20000 metres long.
NTIS uses the historic data to assign each link in the network with a traffic profile.
Traffic profiles contain, for each minute of a day in a given date, the expected travel time for the link to which it is associated. 
These profile values are published together with the actually measured travel times.

Travel time profiles of heavily-used roads vary strongly depending on the time of the day and the day of the week.
However their inter-week variation remains low, with major changes in usage patterns taking longer (order of months) than the biggest temporal unit considered (weeks). 
This slow rate of change with respect to the data frequency illustrates that the calculation of  travel time profiles is a very different task than short-term forecasting.
Since there is no uniformly accepted definition of traffic profiles (typical travel times), here, we will define it as the collection of values that jointly minimise the Root Mean Squared Error (RMSE) or Mean Absolute Relative Error (MARE), when compared with subsequently measured real travel times.

In this paper, we follow up on previous work in \cite{ttprofiles}, providing a novel algorithm capable of generating one week of expected travel times, using a single parameter and using only publicly available data from NTIS.
This approach stems from spectral and statistical analysis of recent historical data and removes the requirement of any prior form of segmentation of times and days into different classes. 
Instead, the algorithm relies on pattern discovery for intra- and inter-day variability which it computes directly from input data.

\subsection{Previous Work} \label{Previous Work}
There is an extensive literature on travel times, although recent research is more focused on shorter term forecasting, with far fewer long term estimation studies \cite{long-term} \cite{long-term-2}. 
Machine learning and statistical analysis methods receive the most attention \cite{should}. 
In machine learning, neural networks are recently having a high relevance \cite{NN} \cite{spectral2}.
Other approaches are closer to the methods here presented: making use of historical data \cite{simple} \cite{dynamic-historic}, differentiating between rush and non-rush hour \cite{peak-historic}, using spectral methods \cite{spectral1} or Locally Weighted Regressions \cite{williams} \cite{sun} \cite{zhong} \cite{chowdhury} \cite{acqua} \cite{vana}.
The Wavelet Transform has been previously found to be useful in combination with Kalman filters \cite{nonlinear}, neural networks \cite{samant} \cite{ghosh} \cite{hojjat} \cite{adeli} and statistical analysis \cite{basu} \cite{hang}; but these either cover short predictions, use the Wavelet analysis of travel times for other purposes such as incident detection, or do not focus on the spectral properties across timescales or are not real-world data based.
Comparisons involving some of these studies are performed in \cite{nikovski}, \cite{lint}, \cite{mori} and \cite{ser}, but they either focus on short-term prediction or do not produce an overall best-performer.
The method here presented is distinct from the previous work mentioned, since it does not use a sample of individual trips, but all conducted over 12 weeks in multiple sites. 
The aggregation period and prediction resolution is minutely, and the prediction horizon is larger.
Among these methods, the transferability to other locations is not often evaluated, being tuned for a specific location with its own specific conditions.
In order to ensure transferability, our method is examined on 39 individual locations across two motorways and independently scored for each site.
The work presented here makes use of a combination of Continuous Wavelet Transform \cite{morletwavelet}, tree decisions, spectral analysis, and Locally Weighted Regression (LWR).

\section{Travel times in Motorways} \label{Travel times in Motorways}
From a user perspective, vehicle travel times are the most relevant measure of the state of the traffic flow on a road link.
The average travel time in a given minute of the day on a given link will be the average time to travel from the entry to the exit loop sensor for all vehicles that passed through.
\begin{figure}[htbp]
\centerline{\includegraphics[width=10cm]{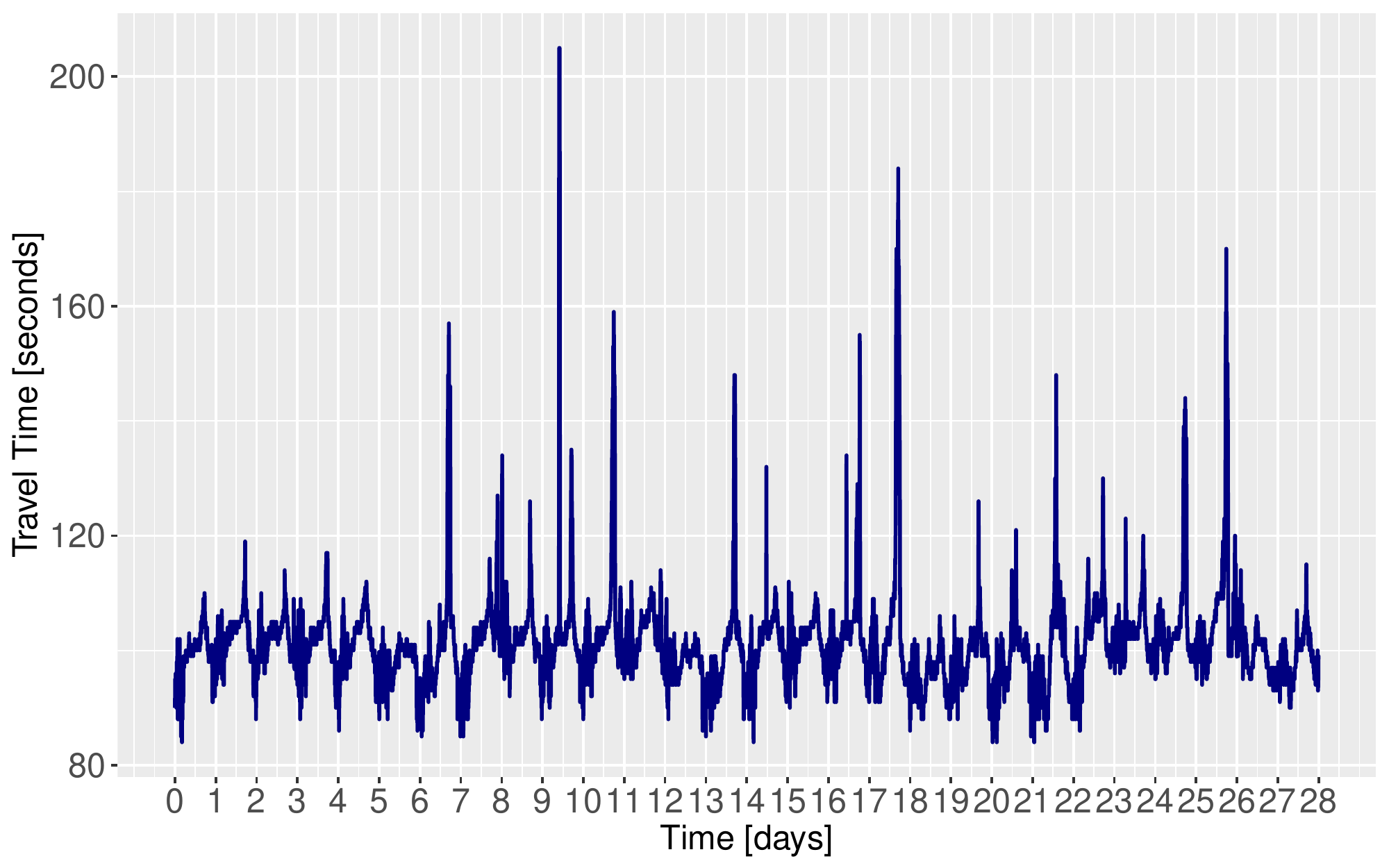}}
\caption{Link 1170079, M6. Travel Time over 28 days of minutely data between 07/Mar/2016 and 04/Apr/2016.}
\label{fig:travel_time}
\end{figure}
From Figure \ref{fig:travel_time}, it can be observed that most of the time, the travel time on a link follows a repeating pattern with minima located at night matching the bounded free flow time (except for speeding drivers).
As the morning rush starts, travel time will rise as traffic jams are generated from the collective drivers' behaviour.
More effects of these collective dynamics can be observed in the afternoon during the evening rush, normally being possible to find a plateau between these two peaks.
Finally, travel times progressively decay towards the night's free-flow regime.
In Figure \ref{fig:travel_time}, we observe a series of spikes found outside of this normally bounded yet oscillating behaviour. 
Travel time in these events can climb up to several times the normal amount.
The predictability in terms of duration and amplitude of these spikes is much lower than the periodic component described above.
\begin{figure}[htbp]
	\centerline{\includegraphics[width=10cm/2]{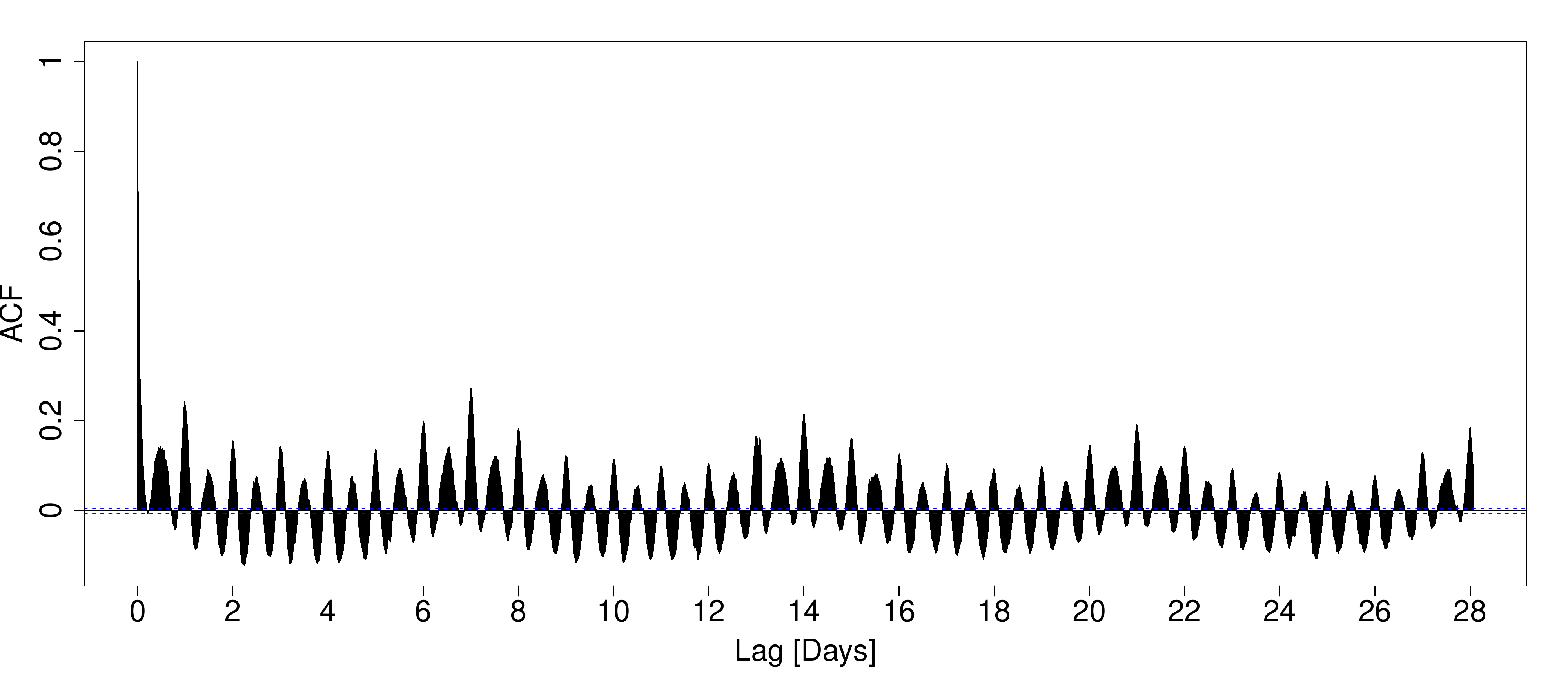}}
	\caption{Autocorrelation function of Travel Time in figure \ref{fig:travel_time} with a maximum lag of 4 weeks (1440 points/day). Double seasonality can be observed in the weekly and daily periods.}
	\label{fig:ACF}
\end{figure}
Lastly, in Figure \ref{fig:ACF}, we can see the Autocorrelation Function for the Travel Time, plotted with a maximum lag of 4 weeks. Here we can see a double seasonal pattern with periods of 1 day and 1 week.
This regularity seen in time travel time series can and is often used by modellers to approximate and forecast travel times as is explored over the next sections.

\section{Basic methods for profile estimation}\label{Basic methods for profile estimation}
\subsection{Exponentially Weighted Moving Average for Profiles} \label{ewma}
As explained in \cite{ttprofiles}, a basic approach to estimating profiles is to apply an Exponentially Weighted Moving Average (EWMA) across a given minute of the available days, assuming that similar behaviour is to be expected at similar times on different days. 
Then, the profile estimation $\hat{x}(i,d+1)$ for the $i-th$ minute of a date $d$, controlling our memory parameter $\alpha \in [0,1]$ to balance the memory of the process and with measured travel time $x_i^d$, will be:
\begin{equation}
\begin{aligned}
\hat{x}^{d+1}_i \!&= \alpha  x^{d}_{i} + (1-\alpha)\hat{x}^{d}_{i} \\ 
&= \alpha  x^{d}_{i} + \alpha^2  x^{d-1}_{i} + (1-\alpha - \alpha^2) \hat{x}^{d-2}_{i}\\ 
&= ...
\end{aligned}
\label{eq:ewma}
\end{equation}
EWMA-based profiles have a main issue: disruptions after large events are generated due to the way in which the memory decays as we can see above in equation \ref{eq:ewma}.
\begin{figure}[htbp]
	\centerline{\includegraphics[width=10cm]{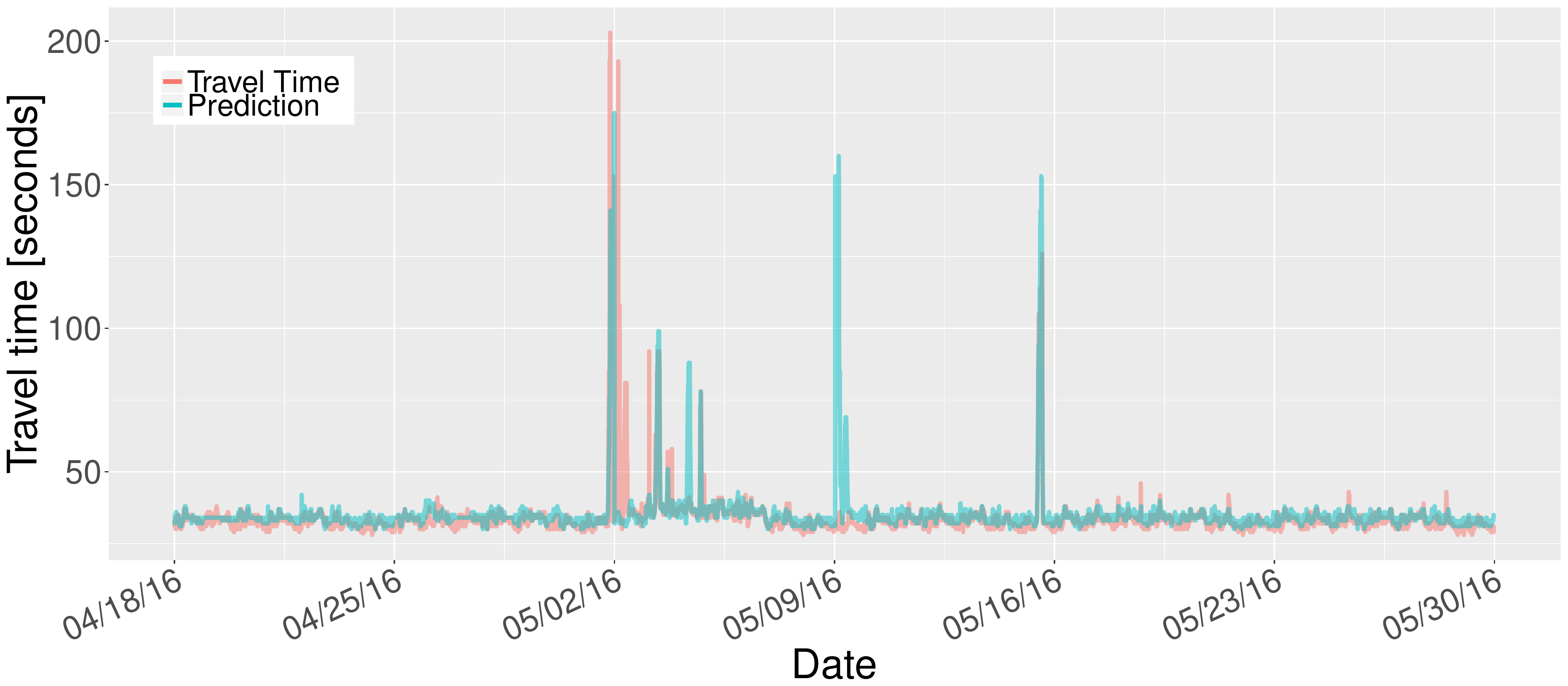}}
	\caption{Skewed EWMA predictions after a large deviation.}
	\label{fig:EWMA}
\end{figure}
Recent measurements receive exponentially greater weights than events farther in the past. 
If a large deviation from the baseline patter occurs, subsequent estimations will have a bias towards partially replicating the deviation and consistently predicting over-estimates until newer data is included and the effect dissipates.

\subsection{Time-based Segmentation}\label{segmentation}
Furthermore, to recognize and use the precise characteristics of specially distinct dates, these methods make heavy use of segmentation based on the date. 
Days are grouped according to pre-defined categories and the EWMA is applied across dates falling in the same category (i.e. Saturdays, weekdays, New Year, ...).
This segmentation in classes, joined with the weakness commented in Section \ref{ewma} can generate lasting perturbations, propagating across weeks into future estimations, yet never being reflected on the travel times measurements. 
A likely instance of such an event is shown in Figure \ref{fig:EWMA}.
Consequently, significant operational and geographical expertise about specific roads is needed to create a valid segmentation, given that the EWMA approach is effective exclusively for recurrent congestion. 
These requirements often lead to the generation of legacy systems which become increasingly difficult to maintain as time passes, their usefulness declining over time or requiring additional efforts for continually training new staff to maintain the system.
These modelling and operational limitations may make segmentation and EWMA based profiles suboptimal both in performance and operation.

\section{Data Selection and Contents}
The M6 and M11 motorways in England are chosen due to their high use and their display of both recurrent and outstanding congestion, being key in several heavily used commuting routes. 
\begin{itemize}
	\item The dataset aggregates 90 days (12 complete weeks) minutely entries (07/03/2016-05/06/2016). This length is chosen to minimise the effect of including other wider partially captured periodicities (e.g. natural seasons) during data gathering.
	\item Links with over 10\% of data missing or containing access ramps were discarded.
	\item The previous condition left 14 different links in the M6 and 25 links in the case of the M11.
	\item Entries missing for 10 minutes or less were linearly interpolated.
	\item Entries missing for over 10 minutes were left as missing values.
\end{itemize}
The algorithm uses 8 weeks of data to predict one entire week ahead.
After a week, the oldest week is deleted, and the most recent one is incorporated, producing a new estimate for the subsequent week.
This procedure is simulated 4 times.
For each link-date pair, the data comprises minutely data, containing: average vehicle travel time in seconds, profile (expected) travel time in seconds, traffic flow in cars/hour and vehicle headway in meters.

\section{Background and Spikes}
\subsection{Characteristics}
As described in \cite{ttprofiles}, if we look at the travel times, we find they operate in two different regimes that we call background and spikes.
The background is stable with small high-frequency fluctuations around a time-varying mean value.
This makes it suitable for seasonal analysis and spectral filtering (smoothing).
In contrast, the spikes are zero most of the time but can quickly climb to extreme values. 
They have much greater amplitude and much lower frequency, creating long reaching effects. 
Although they are non periodic in the time domain, a non-harmonic seasonality contribution associated with recurrent congestion can be extracted via non-parametric regression.

In this context, if we assume Gaussian noise $\xi$, and given the additive properties of wavelet decomposition, the decomposition will be of the form:
\begin{equation}
\textrm{Travel Time}_t  = \textrm{Background}_t + \textrm{Spikes}_t + \xi
\end{equation}
The objective is to separate the signals such that the times of smooth non-congested operation, together with the recurring congestion, are captured in the background and passed through spectral smoothing, mitigating the estimation errors created by the high frequency oscillations and achieving a view of what can be daily observed, so seasonal patterns can be extracted in the shorter and longer periods shown in Fig. \ref{fig:ACF}.
Meanwhile, the spikes, containing the non-recurring congestion, can be searched for any seasonality left on time scales larger than the period in which the travel times oscillate, as also suggested by Fig. \ref{fig:ACF}. 
If performed correctly, the remainder after this seasonal extraction step of the decomposition should contain only isolated events with large deviations from the profile and white noise.
Wavelets are especially well suited for this since they adapt to the relevant scales and location in time, not being a property that is found in a priori explicit forms (splines, polynomials, etc.).
\subsection{Wavelet Time Series Decomposition}\label{decomposition}
\begin{figure*}[htbp]
	\centerline{\includegraphics[width=\linewidth]{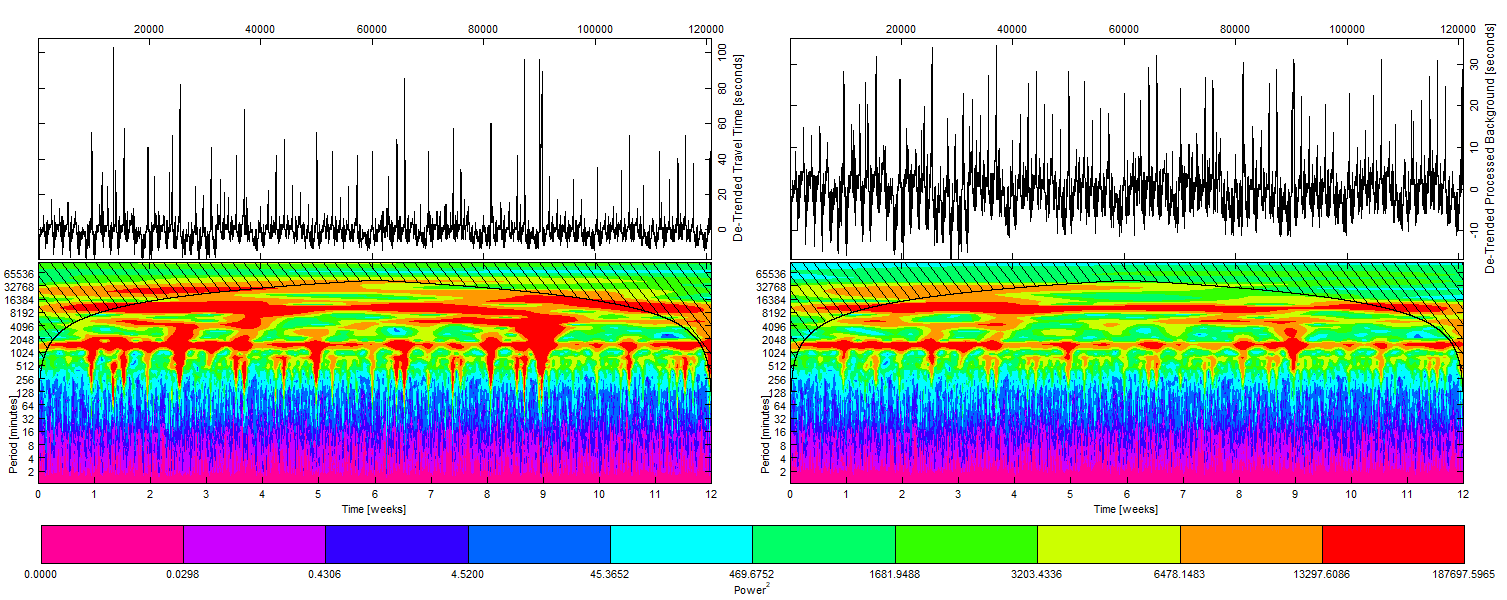}}
	\caption{Left: Power of CWT of the original Travel Time series for Link 1170079 of the M6.
	Right: Power of CWT of the extracted Background $\vec{B_t}$ series, as detailed in Section \ref{decomposition}, for the same link, displaying more stable and periodic dynamics.}
	\label{fig:wt}
\end{figure*}
To perform the decomposition, we will take advantage of the additive properties of the wavelet transform. First the time series $\vec{x_t}$ of length $t$, and elements $x_i$ with $i \in [0,t]$ is turned into a zero mean series $\vec{x_t} - mean(\vec{x_t})$ and used as input for a Continuous Wavelet Transform (CWT) \cite{daubechies} \cite{mallat} using a Morse wavelet \cite{morse} and 140 timescales levels.
The output of this first transform is a complex matrix $\vec{W}_{\textrm{levels}\,\times\, t}$, for the elements of which we calculate their modulus $\rho$, phase $\phi$ and power $P$. 
A heatmap of $P^2$ for the original time series can be seen in the first subplot in Figure \ref{fig:wt}.
In the figure we can observe that the most influential dynamics occurring during the series length (x axis) happen at the timescales (y axis) where it was expected based on Figure \ref{fig:ACF}, namely 1 day (1440 minutes) and 1 week (10080 minutes), based on the $P^2$ values on these bands.
In the figure, we also observe that surges in power across different wavelet timescales occur at the same time as the non-recurring congestion, since in order to approximate this signal, the CWT algorithm needs to combine several wavelets with smaller periods than those dominating the recurring part of the series. 

In order to isolate this non-recurring component, the series is sequentially assessed over all the time domain by taking a horizontal slice for a single timescale level $l$, and generating a series $\vec{x_{t}}^l$.
After fixing $l$, we calculate the Median and Inter Quantile Range of the distribution of values in $\vec{x_{t}}^l$ to search for outliers.
A maximum threshold value is set equal to $T = median(\vec{x_t}) + \alpha * IQR(\vec{x_{t}}^l)$, with $\alpha \in [0,\inf]$ being a parameter that defines how aggressively we target spikes. 
Then, the individual elements of $\vec{x_{t}}^l$, $x_{i}^l$ are individually evaluated: if found below the limit, they are stored in a background container $B_i^l = x_{i}^l$; otherwise, the fraction below the threshold will be nonetheless passed on to the background $\vec{B_i}^l = median(\vec{x_{t}}^l) + \alpha * IQR(\vec{x_{t}}^l)$, with the remaining part going into the spikes storage $\vec{S_i}^l = x_{i}^l - B_{i}^l$.

In the extremes, for $\alpha = 0$ we would find that any deviation from the median is taken into the spikes signal, and with $\alpha = \infty$ only points infinitely distant from the mean would be taken into the spikes.
The results produced here were obtained with $\alpha=1$ to showcase the most basic setup and the effectiveness of even a non-tuned version.

\begin{figure}[htbp]
	\centerline{\includegraphics[width=10cm]{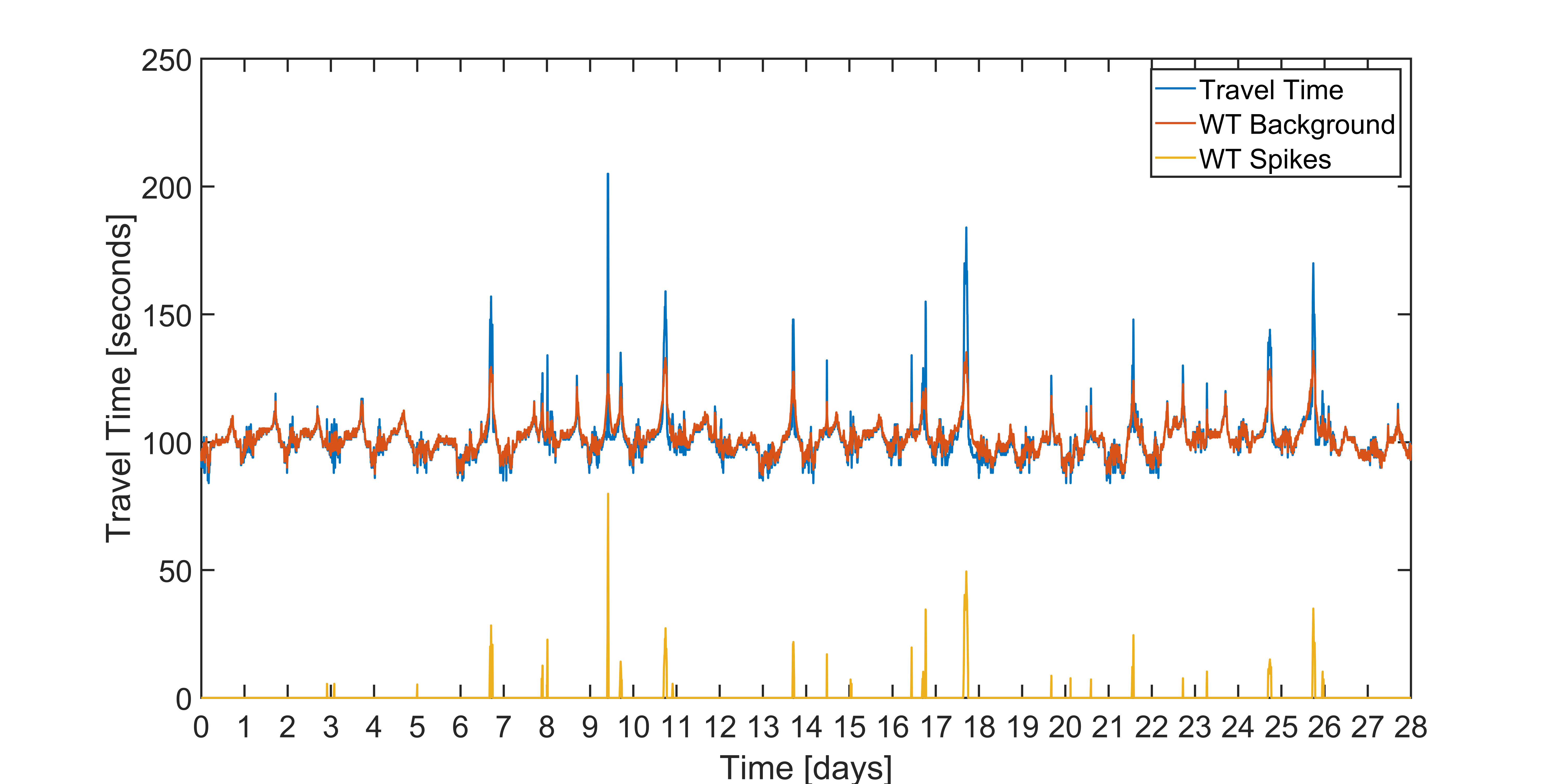}}
	\caption{Recombination post Inverse Wavelet Transform}
	\label{fig:splitting}
\end{figure}

Once all levels have been processed in this manner, we use the previous information about $\phi$ to reconvert the two series from being characterised in terms of $(\rho, \phi)$ to the complex components of the DWT.
After this step, we apply the Inverse DWT to $\vec{B_t}$ and $\vec{S_t}$, obtaining the two series that can be observed in Figure \ref{fig:splitting}.

\subsection{Series recombination  and analysis of background}
The results of the separation of the background can be seen in Fig. \ref{fig:wt}. Here it can be observed that the separated background is still characterised by seasonality dominating the weekly (10080 minutes) and daily (1440 minutes) timescales, and keeping a very similar structure in the regions occupied by the faster dynamics ($<$ 120 minutes) to that of the original series.
Simultaneously, the surges in power across timescales of the original DWT-transformed series (left), which are associated with congestion events, have been greatly reduced or eliminated altogether in the second subplot.

This demonstrates the ability to process a time series showing multiple seasonalities and punctuated by large deviations from the normal dynamics of the process into two separate series, one of which will exclusively contain the baseline dynamics of the process, respecting its structure and natural variability, and the other, which will only contain those large events that occur outside of the smooth operational region.

To reduce noise introduction during the inverse transform, a threshold is applied to $\vec{S_t}$, where elements representing spikes less than 3 seconds in amplitude are set to zero when looking at future estimation.

For future prediction steps, an indicator function is defined for every entry in the series, taking value:
\begin{equation}
    \delta_i^{spike}=
    \begin{cases}
      1, & \text{if } x_i > \text{Threshold}\\
      0, & \text{otherwise.}
    \end{cases}
    \label{delta}
  \end{equation}
  
\section{WARP Travel Time Prediction Algorithm} \label{algorithm}
The objective now is, given the seasonality and separation of the original signal provided above, to generate a time travel prediction model that accounts for the cyclic variations and recurrent congestion but remains resilient to unexpected deviations and rare events.

We aim to provide a robust algorithm that mitigates the propagation of extreme events into the future (unlike EWMA). 
It must work for all locations and not require the use of time segmentation. The trend term must be nearly flat, based on the difference in timescales for the growth of demand on a motorway level and the seasonalities concerning this paper. 
Finally, it must have uncorrelated residuals, Gaussian distributed with mean 0.
Following these requirements, we introduce the WARP algorithm (Wavelet Augmented Regression Profiling) from section \ref{algostart}.

\subsection{WARP: Spectral Component}
\label{algostart}
The background signal shows oscillations of high frequency and low amplitude almost ubiquitously. 
It can be smoothed by discarding, in the frequency domain, the frequencies that in which the oscillations occur and those outside the scope of this study (over 4 weeks and under 4 hours), while keeping those in the information bearing bands by using the Fast Fourier Transform (FFT) \cite{FFT}.
Once this step is performed, and since large events have been removed previously, an EWMA can be applied to the modified weekly power spectra in the FFT, to then compute the Inverse FFT Transform to obtain our background prediction.

\subsection{WARP: Seasonal Component}
The seasonal component is computed via Seasonal-Trend Decomposition based on LOESS (STL) \cite{STL}.
STL is resilient to outliers and can manage any combination of seasonalities, allowing to control their change over time as well as the smoothness of the trend \cite{forecasting}.
We begin by isolating the daily seasonality $S_d$ from the entire background training series using STL.
Trend and remainder are summed and re-analysed for weekly seasonality $S_w$. This step also produces a trend series $T_r$ and a remainder which should be Gaussian distributed, zero mean.
Global seasonality is calculated as $S_g = S_d + S_w$.
We then average the seasonality over the training weeks to obtain a value for each minute of the week we are to estimate.
Then, $T_r$, nearly flat as per \ref{algorithm}, is linearised to obtain a baseline series $B_t$, and the Baseline prediction is $B_p = B_t + G_s$.
Finally, the spikes are searched for any seasonality left on the weekly level $Sp_w$, discarding the trend and remainder terms, and obtaining the final seasonal component as $\textit{SEASONAL} = B_t + {Sp}_w$.

\subsection{WARP Hybrid Profile}
\begin{figure}[htbp]
	\centerline{\includegraphics[width=10cm]{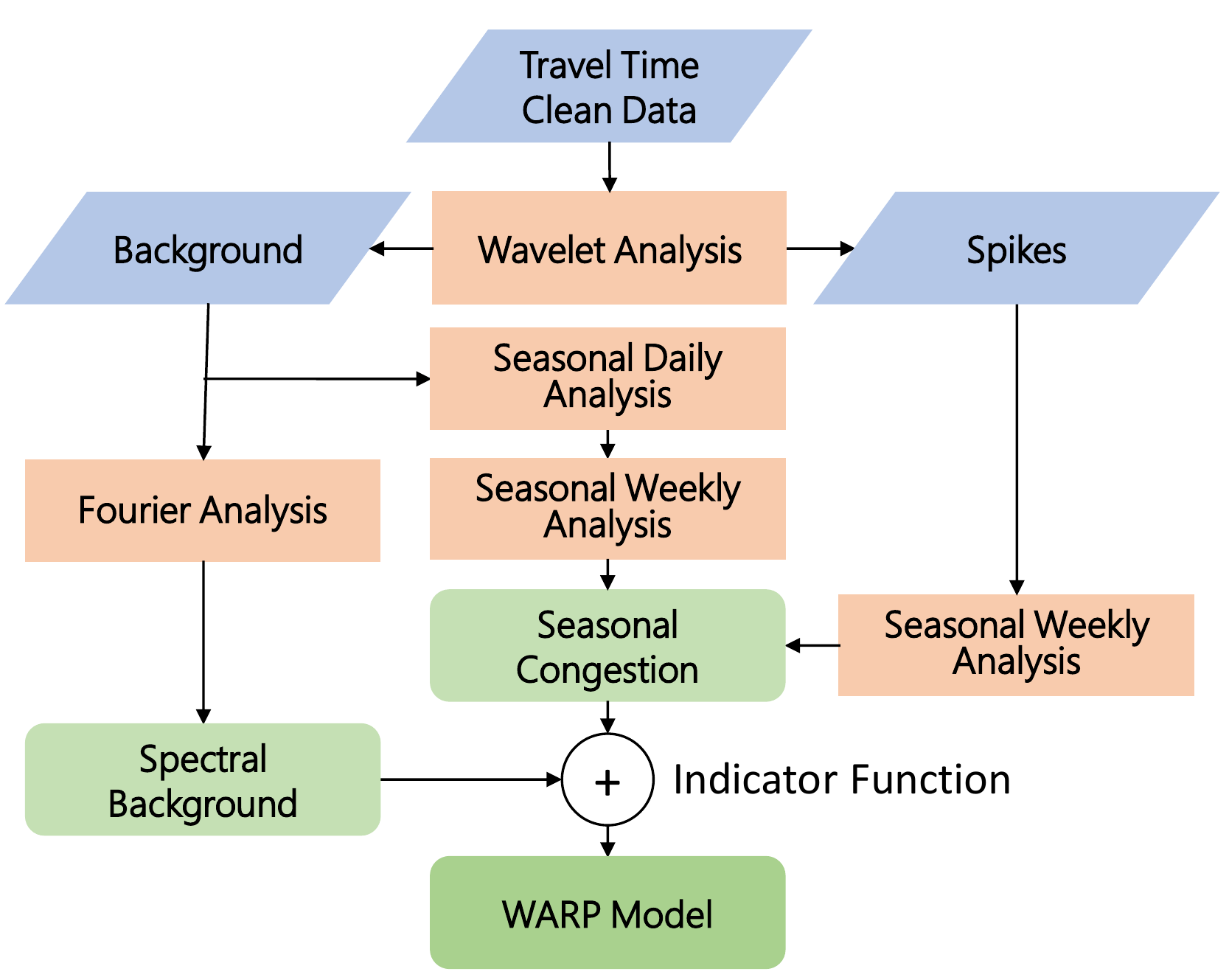}}
	\caption{Flowchart of the data streams in the algorithm}
	\label{fig:dataflow}
\end{figure}
The final WARP profile, containing the spectral and seasonal components, depends on the regime as per Eq. \ref{delta}:
\begin{equation}
\textrm{WARP} = \textrm{Seasonal} * \delta_{\textrm{spike}} + \textrm{Spectral} * (1 - \delta_{\textrm{spike}})
\end{equation}

\section{WARP model validation}

\begin{table*}[bp]
	\caption{MARE Distribution in M6 prediction}
	\centering
	\begin{center}
	\begin{tabular}{SSSSSSSS} \toprule
    {$\textit{Profile / MARE}$} & {$(>-25\%]$} & {$(-25\%,-15\%] $} & {$(-15\%,-5\%]$} & {$(-5\%,5\%)$} & {$[5\%,15\%)$} & {$[15\%,25\%)$} & {$[>25\%)$} \\ \midrule
    \textit{Published M6}  & 1.58 & 0.54 & 5.69 & 56.21 & 30.79 & 3.17 & 2.02 \\
    \textit{Wavelet M6}  & 1.57  & 0.51 & 3.22  & 81.73 & 12.14 & 0.70  & 0.13 \\ \bottomrule
\end{tabular}
	\end{center}
\end{table*}
\begin{table*}[bp]
	\caption{MARE Distribution in M11 prediction}
	\centering
	\begin{center}
	\begin{tabular}{SSSSSSSS} \toprule
    {$\textit{Profile / MARE}$} & {$(>-25\%]$} & {$(-25\%,-15\%] $} & {$(-15\%,-5\%]$} & {$(-5\%,5\%)$} & {$[5\%,15\%)$} & {$[15\%,25\%)$} & {$[>25\%)$} \\ \midrule
    \textit{Published M11}  & 0.85 & 1.15 & 19.21 & 62.83 & 13.21 & 1.65 & 1.10 \\
    \textit{Wavelet M11}  & 0.78  & 0.32 & 3.29  & 81.33 & 12.47 & 1.09  & 0.73 \\ \bottomrule
\end{tabular}
	\end{center}
\end{table*}

In this section we assess the performance of the WARP model in out-of-sample validation and compare it against a null model based on simple segmentation and against the published NTIS profiles (the details of which are not in the public domain.)

\subsection{Simple Segmentation - null model}
Our null model for assessment of the performance of the WARP model is a basic segmentation model that applies uniform weights to the training data in a given time interval from previous weeks. 
On the $i-th$ minute of a week and using the previous $n$ weeks (8 in this paper) as training, the simple segmentation (SS) profile is:
\begin{equation}
\hat{x}(i,n) = \sum_{\textrm{week}=1}^{n} \frac{x^i_n}{n} 
\end{equation}

\subsection{Out-of-sample validation}
Our out-of-sample validation tests compare the WARP profile values against the subsequently measured travel time. 
We choose to use the Mean Absolute Relative Error (MARE) to quantify performance since it allows for a fair comparison of links of different lengths.
The Root Mean Square Error (RMSE) has also been calculated on a motorway level as the average of the RMSE across its links.
\begin{figure}[htbp]
	\centerline{\includegraphics[width=10cm]{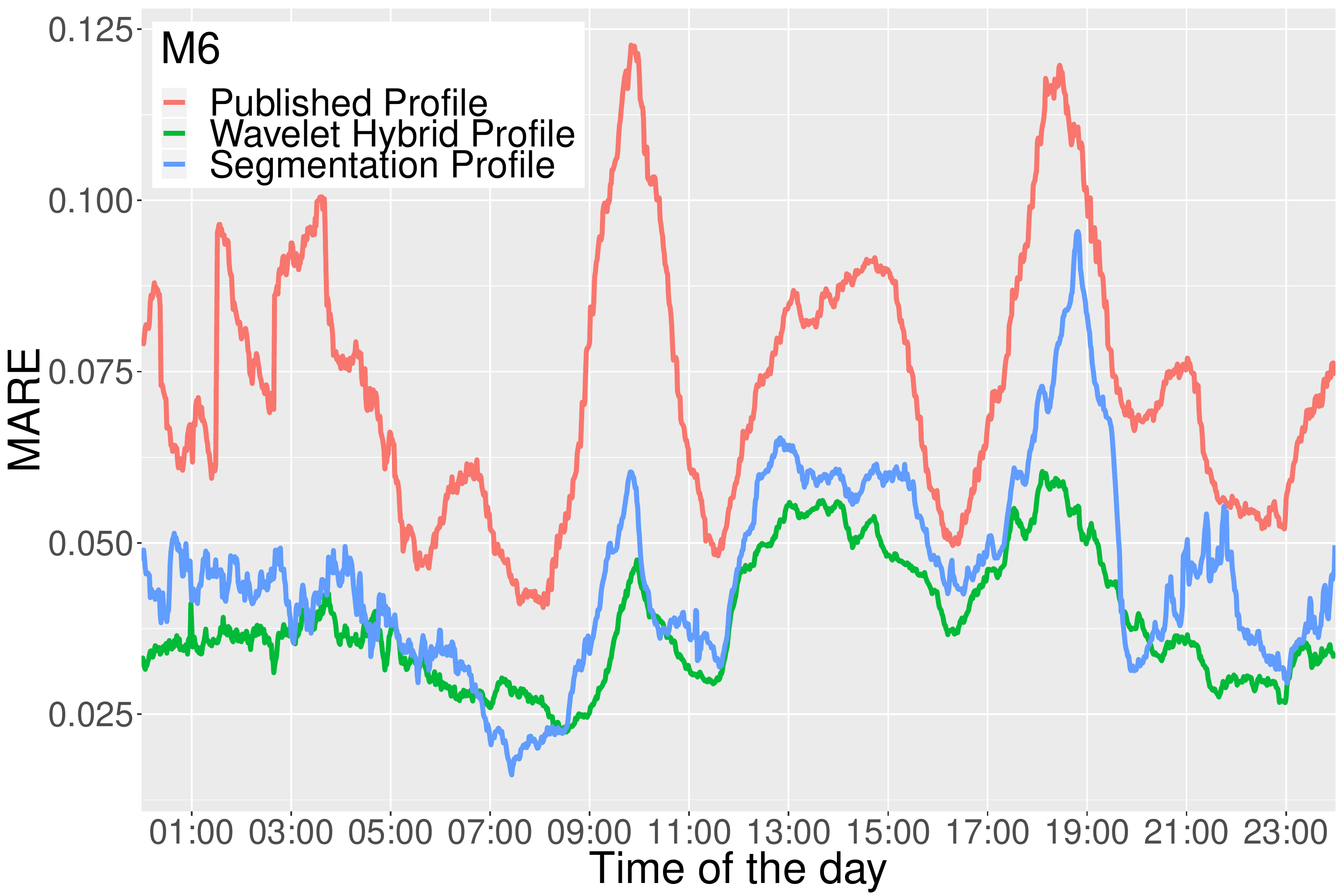}}
	\caption{MARE over the times of the day for the M6}
	\label{fig:m6dt}
\end{figure}
\begin{figure}[htbp]
	\centerline{\includegraphics[width=10cm]{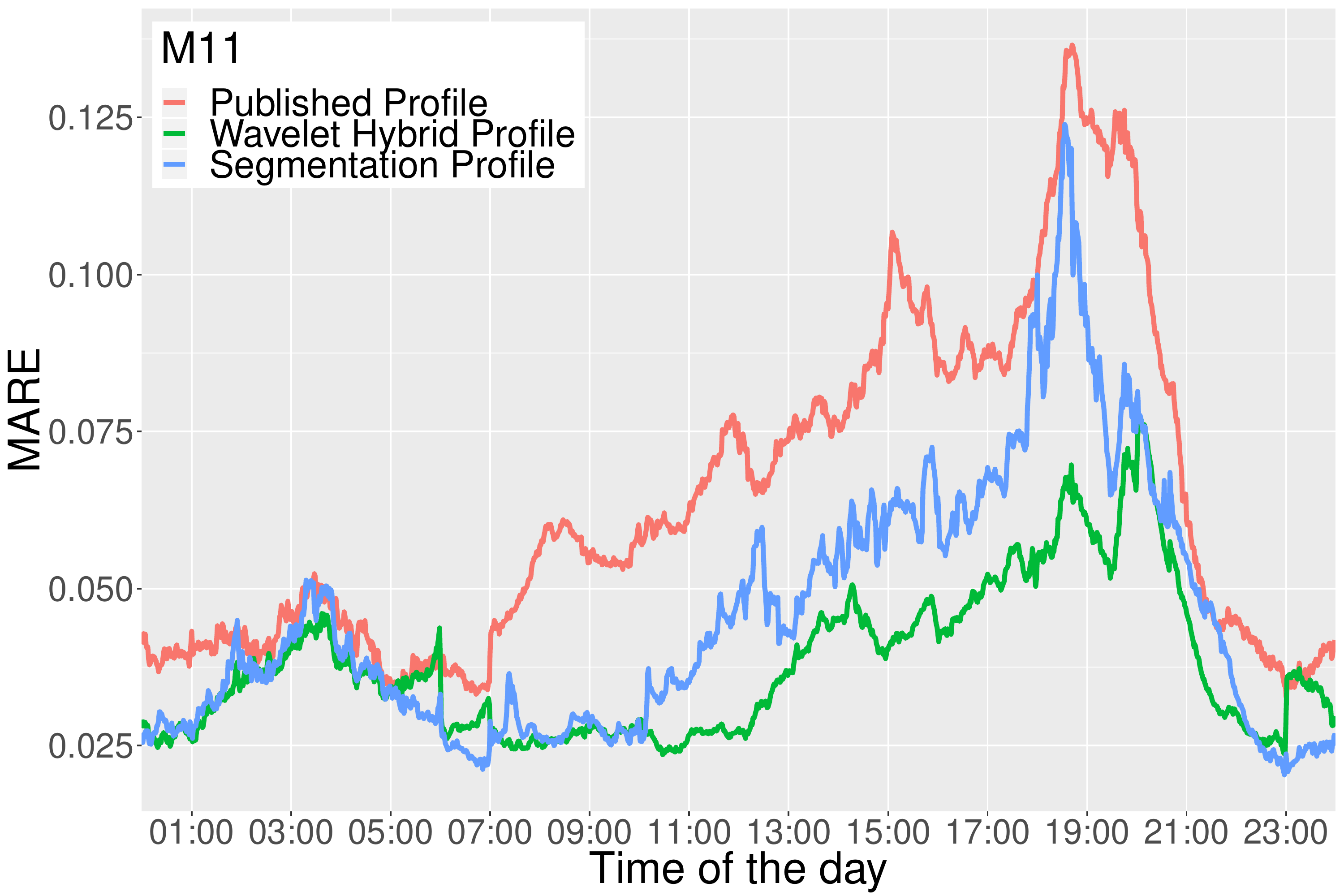}}
	\caption{MARE over the times of the day for the M11}
	\label{fig:m11dt}
\end{figure}
\begin{figure}[htbp]
	\centerline{\includegraphics[width=10cm]{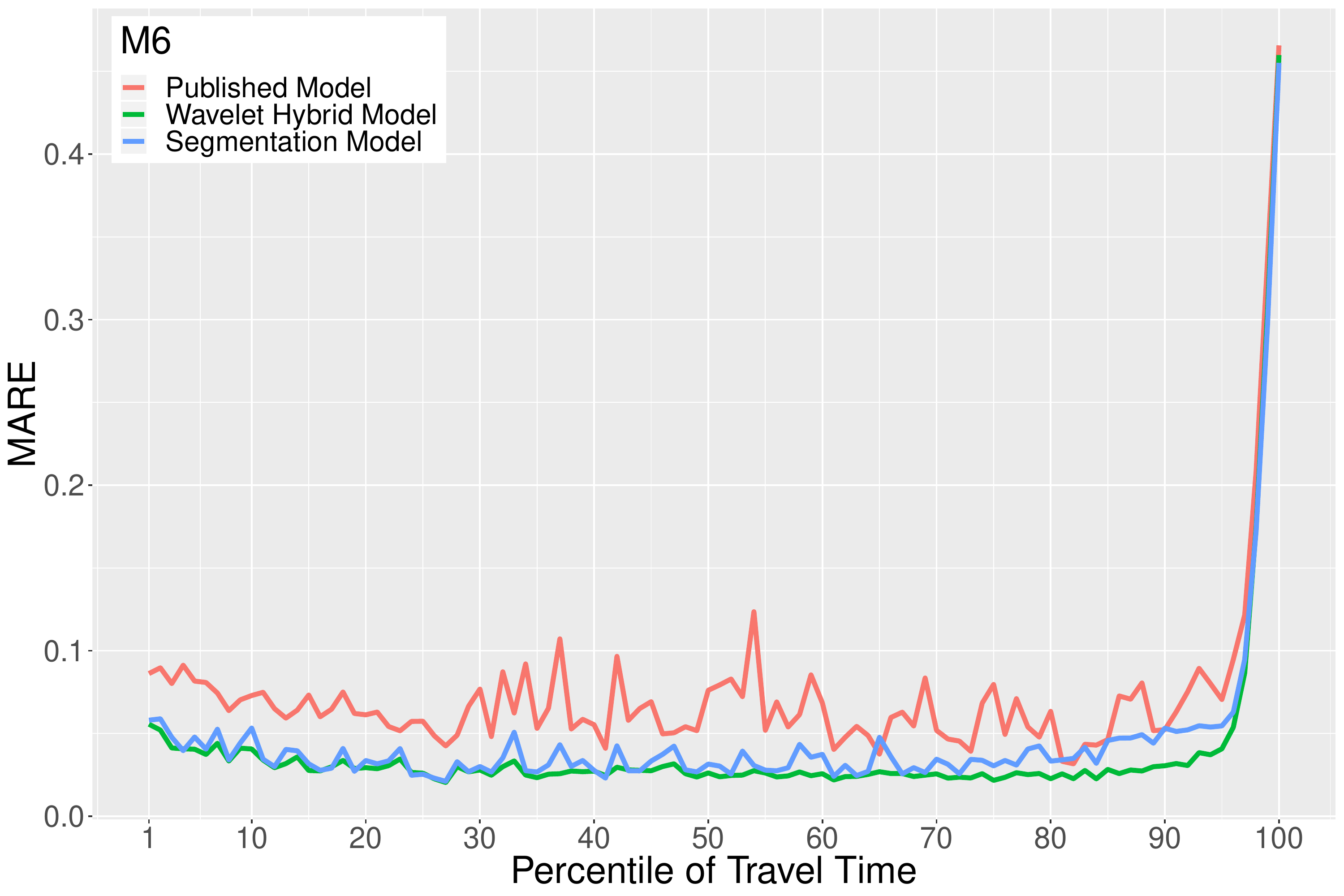}}
	\caption{MARE across percentiles of travel time for the M6.}
	\label{fig:m6q}
\end{figure}
\begin{figure}[htbp]
	\centerline{\includegraphics[width=10cm]{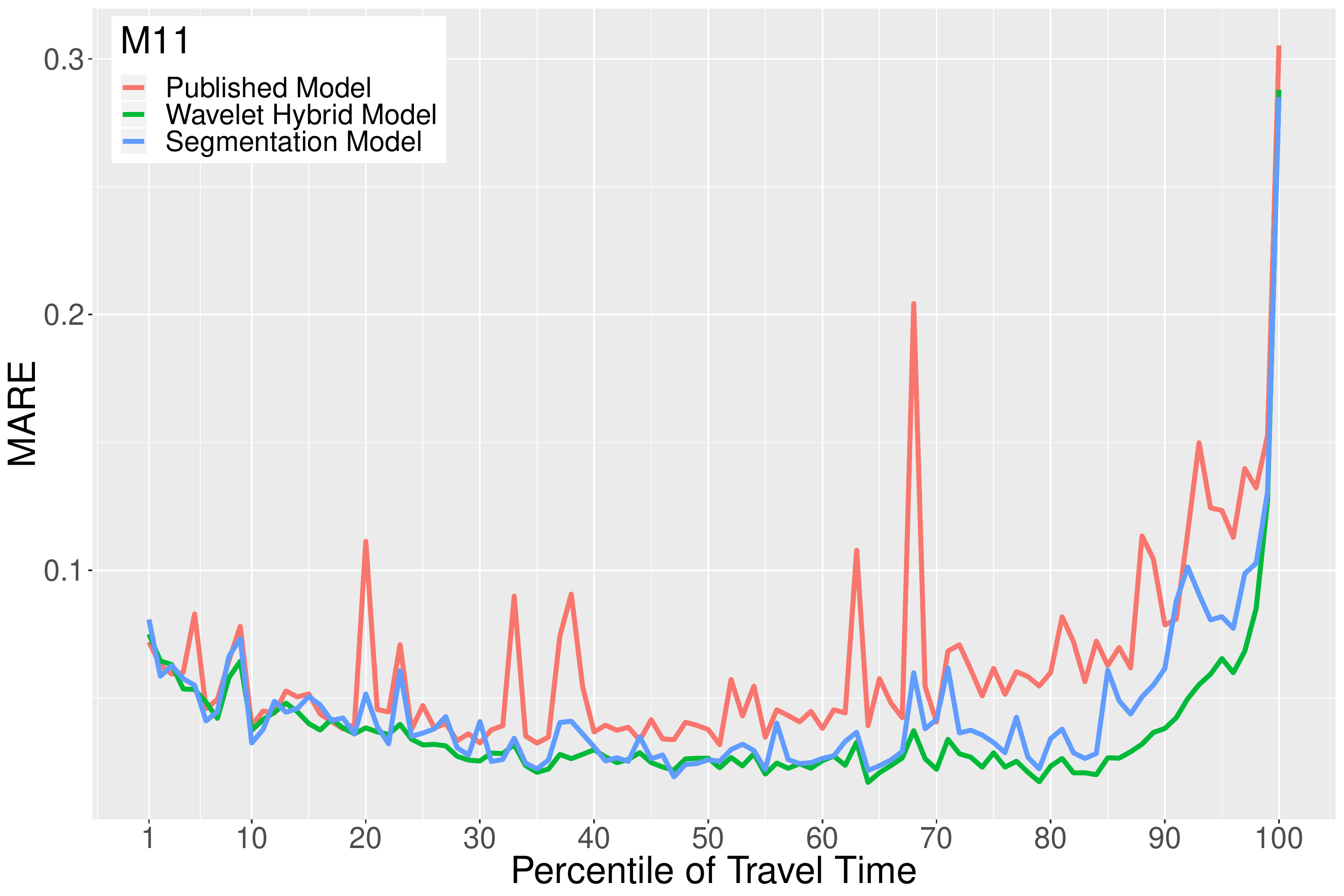}}
	\caption{MARE across percentiles of travel time for the M11.}
	\label{fig:m11q}
\end{figure}

In Figures \ref{fig:m6dt} and \ref{fig:m11dt}, WARP shows lower predictive error than the published profiles and the SS Model for all times and motorways. 
This is most relevant for morning and evening rush hours, where the others' predictive error soars, WARP suffers no meaningful performance worsening relative to the plateau in the middle of the day.
The error at rush hours is reduced by a minimum of 50\% across all cases, reaching as high as 63.4\% in the case of the M6 morning rush.
In Figs. \ref{fig:m6q} and \ref{fig:m11q} it can be observed that the accuracy of the WARP profile is significantly higher than that of the published profiles or the SS Model across nearly all percentiles of travel time.
WARP is most competitive in the upper percentiles of the travel time distribution since it explicitly accounts for the predictable contribution of recurrent congestion to travel time spikes. 
All three models eventually suffer similar errors under the most extreme deviations since these are true outlier and are not amenable to data-driven forecasting.

\begin{table}[htbp]
	\caption{MARE per Link on M6 and M11}
	\begin{center}
		\begin{tabular}{|c|c|}
			\hline
			\textbf{Links M6}&{\textbf{MARE}} \\
			\hline
			117007401& 0.0290\\
			\hline
			117007501& 0.0680\\
			\hline
			117007601& 0.0293\\
			\hline
			117007801& 0.0826\\
			\hline
			117007901& 0.0435\\
			\hline
			117008401& 0.0484\\
			\hline
			117009102& 0.0338\\
			\hline
			117011901& 0.0295\\
			\hline
			117012001& 0.0584\\
			\hline
			117012101& 0.0370\\
			\hline
			117012201& 0.0605\\
			\hline
			117012301& 0.0379\\
			\hline
			117016001& 0.0496\\
			\hline
			123025901& 0.0427\\
			\hline
		\end{tabular}
		\quad
		\begin{tabular}{|c|c|}
			\hline
			\textbf{Links M11}&{\textbf{MARE}} \\
			\hline
			199048301& 0.0510\\
			\hline
			199048701& 0.0279\\
			\hline
			199048801& 0.1053\\
			\hline
			199048901& 0.0239\\
			\hline
			199049002& 0.0306\\
			\hline
			199049101& 0.0213\\
			\hline
			199049402& 0.0223\\
			\hline
			199049501& 0.0181\\
			\hline
			199049702& 0.0297\\
			\hline
			199049801& 0.0408\\
			\hline
			199050002& 0.0445\\
			\hline
			199050101& 0.0272\\
			\hline
			199050202& 0.0288\\
			\hline
			199050901& 0.0433\\
			\hline
			199063301& 0.1203\\
			\hline
			199063701& 0.0500\\
			\hline
			199063801& 0.0265\\
			\hline
			199064203& 0.0230\\
			\hline
			199065202& 0.0259\\
			\hline
			200021668& 0.0280\\
			\hline
			200024801& 0.0233\\
			\hline
			200028639& 0.0241\\
			\hline
			200028641& 0.0188\\
			\hline
			200028645& 0.0435\\
			\hline
			200028648& 0.0499\\
			\hline
		\end{tabular}
		\label{tab1}
	\end{center}
	\label{table:m6mape}
\end{table}
\begin{table}[htbp]
	\caption{Global MARE \& RMSE per Motorway}
	\begin{center}
		\begin{tabular}{|c|c|c|}
			\hline
			\textbf{Motorway}&{\textbf{MARE}}&{\textbf{RMSE [s]}} \\
			\hline
			M6& 0.0385& 0.0464\\
			\hline
			M11& 0.0379& 0.0484\\
			\hline
		\end{tabular}
		\label{mapeglobal}
	\end{center}
\end{table}

\section{Conclusion and Future Work}
This paper has presented an algorithm for separating recurrent and non-recurrent congestion in time travel time series, and generally capable of identifying events following different timescales from those of the baseline dynamic process, and being able to separate them accurately.
More sophisticated analysis of the Wavelet Transform series could be achieved, focusing on the information-bearing bands and using adaptive thresholds for the separation.
Given the flexibility offered by this method, it could be extended for incident detection or adapted to other problems besides road traffic.
The estimation algorithm presented in the last sections meets the requirements described in Section \ref{algorithm} and earlier defined in \cite{ttprofiles}, albeit still using one parameter.
In the future, sensitivity analysis could explore the limits of the algorithm in terms of minimum training data set, as well as maximum performance with increased training.

\bibliographystyle{unsrt}  
%\bibliography{references}  %%% Remove comment to use the external .bib file (using bibtex).
%%% and comment out the ``thebibliography'' section.

%%% Comment out this section when you \bibliography{references} is enabled.

\end{document}